\documentclass[aps,prl,reprint,superscriptaddress,showpacs,amsmath,amssymb]{revtex4-1}

\newcommand{\dd}[1]{\mathrm{d}#1\,}

\renewcommand{\Re}{\mathop{\mathrm{Re}}}
\renewcommand{\Im}{\mathop{\mathrm{Im}}}
\DeclareMathOperator{\Det}{Det}

\DeclareMathOperator{\tr}{tr}

\DeclareMathOperator{\diag}{diag}

\newcommand{\pf}{\mathop{\mathrm{pf}}}
\newcommand{\sgn}{\mathop{\mathrm{sgn}}}

\renewcommand{\vec}[1]{\bm{#1}}

\usepackage{graphicx}
\usepackage{latexsym}
\usepackage{amsmath}
\usepackage{amssymb}
\usepackage{amsfonts}
\usepackage{xcolor}
\usepackage{units}
\usepackage{bm}
\usepackage{verbatim}
\usepackage{hyperref}
\usepackage{esint}
\usepackage{soul}

\definecolor{FG-color}{named}{red}
\definecolor{FSB-color}{named}{magenta}
\definecolor{PV-color}{rgb}{0.97,0.57,0.11}
\definecolor{AB-color}{RGB}{128,0,128}

\definecolor{FG-color2}{named}{red}
\definecolor{FSB-color2}{named}{magenta}
\definecolor{PV-color2}{rgb}{0.97,0.57,0.11}
\definecolor{AB-color2}{RGB}{128,0,128}

\begin{document}

\title{Majorana bound states in hybrid 2D Josephson junctions
  with ferromagnetic insulators
}

\author{P.~Virtanen}
\email{pauli.virtanen@nano.cnr.it}
\affiliation{NEST, Istituto Nanoscienze-CNR and Scuola Normale Superiore, I-56127 Pisa, Italy}

\author{F.~S.~Bergeret}
\email{sebastian\_bergeret@ehu.eus}
\affiliation{Centro de Fisica de Materiales (CFM-MPC), Centro Mixto CSIC-UPV/EHU, Manuel de Lardizabal 4, E-20018 San Sebastian, Spain}
\affiliation{Donostia International Physics Center (DIPC), Manuel de Lardizabal 5, E-20018 San Sebastian, Spain}

\author{E.~Strambini}
\affiliation{NEST, Istituto Nanoscienze-CNR and Scuola Normale Superiore, I-56127 Pisa, Italy}

\author{F.~Giazotto}
\affiliation{NEST, Istituto Nanoscienze-CNR and Scuola Normale Superiore, I-56127 Pisa, Italy}

\author{A.~Braggio}
\email{alessandro.braggio@nano.cnr.it}
\affiliation{NEST, Istituto Nanoscienze-CNR and Scuola Normale Superiore, I-56127 Pisa, Italy}

\begin{abstract}
  We consider a Josephson junction consisting of
  superconductor/ferromagnetic insulator (S/FI) bilayers as electrodes
  which proximizes a nearby 2D electron gas. By starting from a
  generic Josephson hybrid planar setup we present an exhaustive
  analysis of the the interplay between the superconducting and
  magnetic proximity effects and the conditions under which the
  structure undergoes transitions to a non-trivial topological phase.
  We address the 2D bound state problem using a 
  general transfer matrix
  approach that reduces the problem to an effective 1D Hamiltonian. This allows
  for straightforward study of topological properties in different
  symmetry classes. As an example we consider a narrow channel coupled
  with multiple ferromagnetic superconducting fingers, and discuss how
  the Majorana bound states can be spatially controlled by tuning the
  superconducting phases. Following our approach we also show the energy spectrum, the free energy and finally the multiterminal Josephson current of the setup.
\end{abstract}

\maketitle

\emph{Introduction}. Majorana bound states (MBS) \cite{kitaev2001-umf} have
been proposed as a building block for solid-state topological quantum
computation \cite{Kitaev2003,*Nayak2008}.  Different setups have been
discussed theoretically  
\cite{Fu2008,oreg2010-hla,*stanescu2010-pea,*stanescu2011,*sau2010-gnp,*Lutchyn2010,alicea2010-mfi,*Alicea2012,
  hell2017-tdp,*hell2017-cbm,sticlet2017,pientka2017-tsp,prada2012,rainis2013}
--- many of them relying on the combination of materials with strong
spin-orbit coupling, superconductors and external magnetic field.
Following these suggestions, experimental research has been focused on
hybrid structures between semiconducting nanowires \cite{Mourik2012}
and more recently two-dimensional electron systems
\cite{Tanaka2009,Williams2012,OostingaPRX2013,*Sochnikov2015,Kurter2015,kjaergaard2016quantized,Bocquillon2016,*Wiedenmann2016,shabani2016two,*zutic2016,suominen2017scalable}
in proximity to superconducting leads. Setups based on 2DEGs are of
special interest, as they benefit from the precise control of the 2DEG
quantum well technology developed in the last 40 years.  Ideally, the
external magnetic field should act as a pure homogeneous Zeeman field
acting on the conduction electrodes of the semiconductor. In practice, in the presence of superconductors,
the situation is more complex due to orbital effects and spatial inhomogeneity
due to magnetic focusing \cite{Lee2013,paajaste2015,*tiira2017magnetically,zuo2017}.
 
\begin{figure}
  \includegraphics{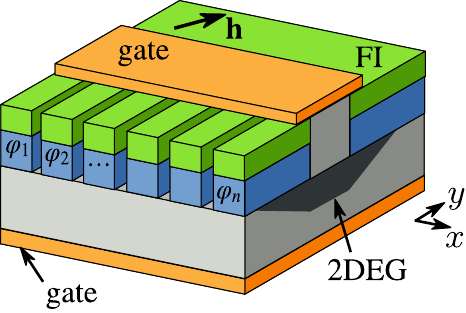}
  \caption{\label{fig:setup}
    Schematic of a narrow semiconductor channel (2DEG/nanowire) contacted to
    ferromagnetic insulator--superconductor bilayers. Phase
    differences $\varphi_j$ of the superconducting $j$-th fingers are imposed across the semiconductor channel. Multiple fingers can be used to precisely control the position of the topological bound states in the junction.}
\end{figure}

The aim of the present work is twofold: to propose a setup for
hosting and manipulating MBS at zero external magnetic field, and to
discuss a general analytical approach for describing the
transport and topological properties of 1D boundary systems with
generic symmetries.  The proposed setup is sketched in
Fig.~\ref{fig:setup} which consists of a
2DEG\cite{Deon11,*Deontwo11,*Amado13,*Amadotwo2013,*Fornieri13}
coupled to ferromagnetic insulator/superconductor (FI/S) electrodes.
Related 2D
systems have been recently explored in
Refs.~\cite{pientka2017-tsp,hell2017-tdp,sticlet2017,konschelle2016,*konschelle2016b}.
The magnetic proximity effect from the FI induces an effective
exchange field $\vec{h}$ in the superconductors breaking time-reversal
symmetry and resulting in a spin splitting of the density of states.
Experimentally, manufacturing S/FI films is well demonstrated.
\cite{moodera1988electron,*hao1990spin,Strambini2017}.
We approach the theoretical problem by developing an exact method
that provides a systematic dimensional reduction procedure based
on a continuum transfer matrix approach
\cite{lee1981,hatsugai1993-esi,mora1985,garciamoliner1990-gtm}, which
in certain aspects is closely related to scattering theory
\cite{beenakker1991-jct,*beenakker2015-rmt}.
The effective 1D boundary Hamiltonian obtained provides access to the
energy spectrum, the free energy
\cite{gelfand1960-ifs,forman1987-fdg,waxman1994,kosztin1998-feo},
and the multiterminal Josephson currents in the
  setup. An analytically tractable 1D topological invariant
  also emerges in a natural
  manner.  The approach also applies to 2DEG strongly coupled to superconductors
  via transparent interfaces, required for large topological energy gaps.
Here, we apply this method for our class D problem
\cite{ryu2010-tia,*RMPChiu2016} and discuss phase-controlled
manipulation \cite{Fu2008,romito2012} of the MBS with inhomogeneous
multiple S/FI "fingers".

\emph{Model.}
To model the junction of Fig.~\ref{fig:setup} we use the Bogoliubov--de-Gennes Hamiltonian of the proximized 2DEG
in the basis
$(\psi_\uparrow,\psi_\downarrow,-\psi_\downarrow^\dagger,\psi_\uparrow^\dagger)$,
\begin{align}
  \label{eq:BdGe}
  \mathcal{H}
  =
  \hat{\nabla}
  \cdot
  \frac{-\tau_3}{2m(x,y)}
  \hat{\nabla}
  -
  \mu(x,y)\tau_3
  -
  \vec{h}(x,y)\cdot\vec{\sigma}
  +
  \hat{\Delta}(x,y)
  \,.
\end{align}
Here,
$\hat{\nabla} = \nabla +
i\hat{x}[\vec{\alpha}_x(x,y)\cdot\vec{\sigma}] +
i\hat{y}[\vec{\alpha}_y(x,y)\cdot\vec{\sigma}]=\hat{x}\hat{\partial}_x+\hat{y}\hat{\partial}_y$
contains the spin-orbit su(2) vector potential. \cite{froelich1993}
The superconducting order parameter is
$\hat{\Delta}=\frac{1}{2}[\tau_1+i\tau_2]\Delta+\frac{1}{2}[\tau_1-i\tau_2]\Delta^\dagger$,
$\vec{h}$ is the exchange field, and $m$ the effective mass.
Moreover, $\tau_{1,2,3}$ and $\sigma_{x,y,z}$ are Pauli matrices in
the Nambu and spin spaces, respectively. We assume that inside the
``lead'' region, $|y|>L/2$, the parameters are independent of $y$ but
may vary along $x$.

\emph{Reduction to 1D.}  To study the Andreev bound
states (ABS) localized in the  $|y|<L/2$ region, we reduce the problem from 2D to 1D using the transfer matrix \cite{lee1981,hatsugai1993-esi,mora1985,garciamoliner1990-gtm,waxman1994,titov2006-jeb}.
Consider first the 2D Schr\"odinger equation
$G^{-1}(\epsilon)\psi=[\epsilon-\mathcal{H}]\psi=0$ and   define the vector $\mathbf{u}=(\psi; \tau_3(2m)^{-1}\hat{\partial}_y\psi)$. 
$\psi$ satisfies the differential equation when 
$\partial_y\mathbf{u}(y)=\mathbf{W}(y)\mathbf{u}(y)$, where
\begin{gather}
  \label{eq:transfermatrix}
  \mathbf{W}(x,y) = \begin{pmatrix}
    -i\vec{\alpha}_y(x,y)\cdot\vec{\sigma}
    &
    2m(x,y)\tau_3
    \\
    \mathcal{H}\rvert_{\hat{\partial}_y=0}-\epsilon
    &
    -i\vec{\alpha}_y(x,y)\cdot\vec{\sigma}
  \end{pmatrix}
  \,.
\end{gather}
The fundamental matrix $\mathbf{\Psi}(y,y')$, such that
$\mathbf{u}(y)=\mathbf{\Psi}(y,y')\mathbf{u}(y')$, satisfies
$\partial_y\mathbf{\Psi}(y,y')=\mathbf{W}(y)\mathbf{\Psi}(y,y')$ with
$\mathbf{\Psi}(y,y)=\mathbf{1}$.  Below we denote Pauli matrices in
the above $2\times2$ space with $\gamma_{1,2,3}$.  Note that
$\mathbf{W}$ and $\mathbf{\Psi}$ are operators in $x$-basis, and
in a uniform system,
$\det[\mathbf{W}(k_x) - ik_y
\mathbf{1}]=(2m)^4\det[\mathcal{H}(k_x,k_y)-\epsilon]$.  At the
interfaces with the leads, $y=\pm{}L/2$, $\psi$ satisfies boundary
conditions of the form $\psi+A_\pm\hat{\partial}_y\psi=0$. The
coefficients $A_\pm$ contain information about the FI/S leads and are
determined by their $\mathbf{W}$ matrices.  The boundary conditions
can be expressed as
$\mathbf{M}_{y}\mathbf{u}(y)\equiv{}[\mathbf{P}_-\mathbf{\Psi}(-L/2,y)+\mathbf{P}_+\mathbf{\Psi}(L/2,y)]\mathbf{u}(y)=0$,
where
\begin{align}
  \label{eq:P}
  \mathbf{P}_- &= \begin{pmatrix}
    1 & 2mA_-\tau_3
    \\
    0 & 0
  \end{pmatrix}
  \,,
  &
  \mathbf{P}_+ &= \begin{pmatrix}
    0 & 0
    \\
    (2m)^{-1}\tau_3A_+^{-1} & 1
  \end{pmatrix}
  \,.
\end{align}
The bound state energies are then determined by
$\Det \mathbf{M}_{y}(\epsilon)=0$
where $\Det$ is the (functional) determinant in the matrix and $x$ spaces.
It is independent of $y$ because $\Det\mathbf{\Psi}(y,y')=1$.
We can characterize the ABS with a 1D boundary/defect Hamiltonian
\cite{lee1981,peng2017} based on the Green function $G$:
\begin{align}
  \label{eq:Hydef}
  H_y = \epsilon - L^{-1}G(y,y)^{-1}
  \,.
\end{align}
The transformation~\eqref{eq:transfermatrix} provides an explicit connection
between $G$ and $\mathbf{M}_y$, typical \cite{kosztin1998-feo} for such differential
equation systems: \cite{epaps}
\begin{align}
  \label{eq:Hyexpr}
  H_y
  &=
  \epsilon
  -
  2L^{-1}
  [
  \mathbf{M}_y^{-1}
  \gamma_3
  \mathbf{M}_y
  ]_{12}^{-1}
  \,,
  \\
  \label{eq:detHy}
  \Det(\epsilon - H_y)
  &=
  \Det{}(\mathbf{M}_y)
  \Det^{-1}([\frac{L}{2}\mathbf{P}_-\mathbf{\Psi}_-]_{12}[\mathbf{P}_+\mathbf{\Psi}_+]_{22})
  \,,
\end{align}
where $\mathbf{\Psi}_\pm=\mathbf{\Psi}(\pm\frac{L}{2},y)$ and
$[\mathbf{X}]_{12}=\begin{pmatrix}1&0\end{pmatrix}\mathbf{X}\begin{pmatrix}0&1\end{pmatrix}^T$,
$[\mathbf{X}]_{22}=\begin{pmatrix}0&1\end{pmatrix}\mathbf{X}\begin{pmatrix}0&1\end{pmatrix}^T$.
Solutions to the eigenproblem $H_y(\epsilon)\psi=\epsilon\psi$
give the bound state energies.
Zeros of the denominator of Eq.~\eqref{eq:detHy} correspond to a bound
state in either half of the system cut into two with a hard-wall
boundary condition $\psi(y)=0$.
If the leads are topologically trivial so that the edge of the cut system
is gapped, $H_y$ is typically nonsingular at low energies.

\emph{Topological order.} Hamiltonian \eqref{eq:BdGe} has the
charge-conjugation symmetry $\mathcal{H}=-C^\dagger\mathcal{H}^*C$,
$C=-\sigma_y\tau_y$. As
$H_y$ inherits the symmetry of $\mathcal{H}$, its topological
properties can be characterized by the low-energy part of the
1D bulk invariant of class D:~\cite{kitaev2001-umf}
\begin{align}
  \chi(y) = -\sgn\pf{} CH_y(\epsilon=0,k_x=0)
  \,,
\end{align}
where $\pf$ is the Pfaffian of a $4\times 4$ matrix.
Note that $\chi(y)$ can change sign only if bound states cross
$\epsilon=0$, or when $H_y$ has a singularity there.\footnote{The
  singularities of $H_y$ are independent of the lead order parameter
  phase, and do not generically occur at $\epsilon=0$, $k_x=0$ in
  narrow channels for the case considered here [cf. Eq.~\eqref{eq:sigma} and
  \cite{sticlet2017}].} As a consequence, zero-energy bound states
are expected between regions with different $\chi$. This argument can be
generalized to other symmetry classes and formally also to 0D invariants
in systems of finite size along $x$.

\begin{figure}
  \includegraphics{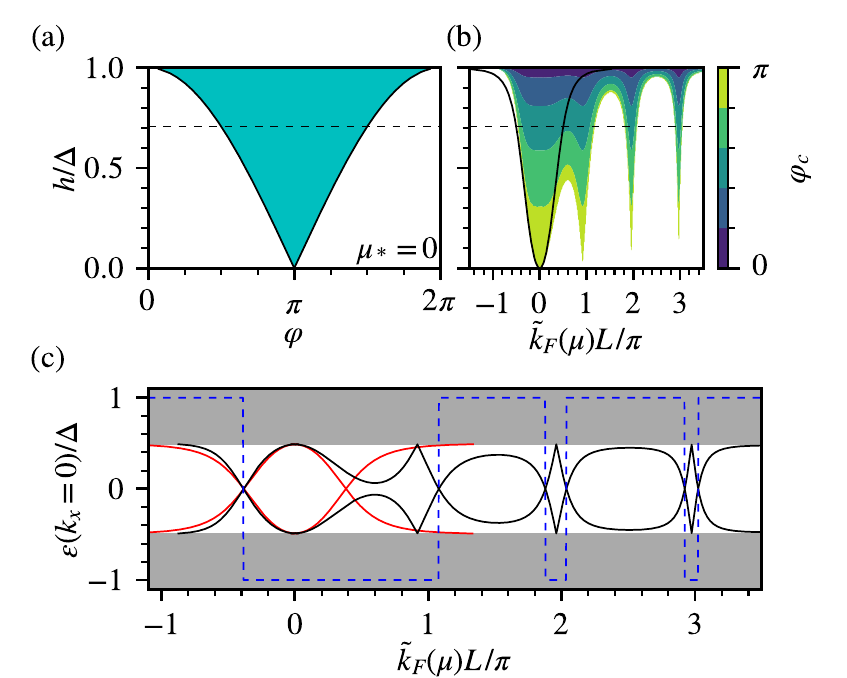}
  \caption{\label{fig:phasediagram}
    Topological phase diagram.
    (a)
    The quantity $\chi(y=0)$ (shaded: $-1$) and the 1D narrow-channel invariant
    $\mathcal{M}$
    (solid line: sign change),
    for $\vec{\alpha}_x\perp\vec{h}$, $\vec{\alpha}_y=0$.
    Dotted line indicates the Chandrasekhar--Clogston limit $h=\Delta/\sqrt{2}$.
    (b)
    Threshold $\varphi_c$, such that $\chi=-1$ for $\varphi_c<\varphi<2\pi-\varphi_c$,
    for varying $\mu$. Here, $\tilde{k}_F(\mu)\equiv\sqrt{2m|\mu_*|}\sgn\mu_*$.
    The threshold for nontrivial state at $\varphi=\pi$ in the narrow-channel 1D model
    is also shown (solid line).
    (c)
    Mode spectrum at $k_x=0$, exact (solid black) and from $\mathcal{H}_{\rm eff}$ (solid red),
    and $\chi$ (dashed), for $h/\Delta=0.5$, $\varphi=\pi$.
    We set $k_S=5/\xi$, $\vec{\alpha}_x=\hat{x}0.5/\xi$, $\vec{\alpha}_y=0$, $L=0.25\xi$, $k_x=0$,
    and $\xi=\hbar/\sqrt{2m|\Delta|}$.
  }
\end{figure}

\emph{Infinite superconducting leads.} For this specific case, we determine
the $A_\pm$ factors from the $\mathbf{W}$ matrices in the
leads, $\mathbf{W}_\pm$. We define
the projectors $\tilde{\mathbf{P}}_{+(-)}=\frac{1+(-)\sgn{}\mathbf{W}_{+(-)}}{2}$ to
growing(decreasing) modes with momenta $+(-)\Im{}k_y<0$ in the
upper(lower) lead, where $\sgn$ is the matrix sign function.  The mode
matching conditions for bound states can then be written as
$\tilde{\mathbf{P}}_\pm\mathbf{u}(\pm{}L/2)=0$.  At energies where the
spectrum of the leads is gapped, half of the modes are growing and
half are decreasing.
Since $\tilde{\mathbf{P}}_\pm$ are then half-rank
matrices, one can generally find $\mathbf{R}$ such that
$\tilde{\mathbf{P}}_\pm=\mathbf{R}\mathbf{P}_\pm$ where
$\mathbf{P}_\pm$ have the structure of Eq.~\eqref{eq:P}.
For $\mu_{\rm lead}\to\infty$ and no spin-orbit interaction in the
leads, direct calculation gives \cite{epaps}
$A_\pm=\mp\frac{1}{ik_S}\tau_3g\tau_3$, where $g$ is the
quasiclassical \cite{eilenberger1968-tog} Green function
$g=[(\epsilon-\vec{h}\cdot\vec{\sigma})\tau_3+\Delta\tau_1]/\sqrt{(\epsilon-\vec{h}\cdot\vec{\sigma})^2-|\Delta|^2}$
and $k_S=\frac{m}{m_{\rm lead}}\sqrt{2m_{\rm lead}\mu_{\rm lead}}$ a
mismatch parameter.

\emph{Narrow-channel expansion.}  Consider now a narrow channel with a
Hamiltonian constant on $|y|<L/2$. Then
$\mathbf{\Psi}(y,y')=e^{(y-y')\mathbf{W}}$. Expanding to leading orders in $L$
($\lesssim{}[2m(|\mu|+|\epsilon|)]^{-1/2}$) in Eq.~\eqref{eq:Hyexpr}
we obtain
\begin{equation}
  \label{eq:sigma}
  \begin{split}
  \mathcal{H}_{\rm eff}
  &=
  H_{y=0}
  =
  \mathcal{H}\rvert_{\hat{\partial}_y=0}
  +
  \Sigma
  \,,
  \\
  \Sigma
  &=
  -
  \frac{\tau_3}{2mL}(A_-^{-1} - A_+^{-1})
  -
  \frac{\tau_3}{4m}(A_-^{-2} + A_+^{-2})
  \\
  &\quad
  +
  \frac{\tau_3}{4m}[A_-^{-1}+A_+^{-1},-i\vec{\alpha}_y\cdot\vec{\sigma}]
  +
  \mathcal{O}(L^1)
  \,.
  \end{split}
\end{equation}
When $\vec{\alpha}_y$ is parallel to the exchange field of
the leads, it commutes with $A_{\pm}$ and does not contribute.
  Imperfections in the S/N interfaces may also be included in this model
  and will affect the precise form of $A_\pm$.
  The above Hamiltonian is obtained
  via operator manipulations, and we did not need to e.g.
  select a variational wave function basis.

Within the  quasiclassical limit in the leads and by setting for simplicity $\vec{\alpha}_y\parallel\vec{h}$ , Eq.~\eqref{eq:sigma}
can be written as
\begin{align}
\label{Heffqc}
  \mathcal{H}_{\rm eff}(\epsilon)
  &=
  \Bigl[
  -
  \frac{1}{2m}
  (\partial_x + i\vec{\alpha}_x\cdot\vec{\sigma})^2
  -
  \mu
  +
  \frac{k_S^2}{2m}
  \Bigr]
  \tau_3
  \\
  \notag
  &\quad
  -
  \vec{h}_{*}(\epsilon)\cdot\vec{\sigma}
  +
  \hat{\Delta}_{*}(\epsilon)
  -
  (Z_*^{-1}(\epsilon)-1)\epsilon
  \,
\end{align}
yielding an effective 1D Hamiltonian with an energy dependent order
parameter $\hat{\Delta}_*$, an exchange field $\vec{h}_*$, a potential
shift, and an energy renormalization [see Eqs.~(S6) in \cite{epaps}
for explicit expressions]. We have neglected the $k_x$ dependence of
$A_\pm$, by assuming $k_x\ll{}k_{F,\mathrm{lead}}$.
 For leads with identical
$|\Delta|$, $|\vec{h}|$ and phase difference $\varphi$,
we find at $\epsilon\to0$,
$Z_*^{-1}=1+D_*|\Delta|^2/(|\Delta|^2-|\vec{h}|^2)$ and
$\Delta_*=D_*|\Delta|\cos\frac{\varphi}{2}$,
$\vec{h}_*=D_*\vec{h}$,
where $D_*=2\hbar^2k_S/[2mL\sqrt{\Delta^2-|\vec{h}|^2}]$.
At low energies, Eq.~\eqref{Heffqc} is similar to widely studied  
quantum wire models \cite{oreg2010-hla,stanescu2010-pea}, and characterized by
the same 1D topological invariant in class D \cite{kitaev2001-umf}
$\mathcal{M}=\sgn\pf C\mathcal{H}_{\rm eff}(k_x=0)\pf
C\mathcal{H}_{\rm eff}(k_x=\infty)$.

The superconducting self-energy in Eq.~\eqref{Heffqc} in the limit
considered here ($L\to0$, transparent NS interfaces) turns out to be
similar in form to weak-coupling tunneling models,
\cite{stanescu2010-pea,sau2010-gnp,alicea2010-mfi,sau2011-pna,stanescu2013}
derived projecting onto lowest quantum well confined modes in the
N-region. The explicit expressions for the prefactors, the shift in
the potential, and $\alpha_y$ spin-orbit obtained here are not found
in typical tunneling approaches.  The magnetic proximity effect from
ferromagnetic superconductors that affects the energy dependence of
both the superconducting and exchange self-energies, on the other hand
in principle can be captured also by a tunneling calculation.

\emph{Phase diagram and spectrum.}  Figure~\ref{fig:phasediagram}(a)
shows $\chi$ and the 1D invariant
$\mathcal{M}=\sgn[\mu_*^2+|\Delta_*|^2-h_*^2]$, where
$\mu_*=\mu-\frac{k_S^2+\alpha_x^2}{2m}$, for a class D narrow
Josephson junction, translationally invariant along $x$,
under a phase difference $\varphi$. This is in agreement with the
phase diagram presented in
Ref.~\onlinecite{pientka2017-tsp}, for the specific value of $\mu_*$.\footnote{Phase dependent
  topological phase diagrams are observed also
  elsewhere\cite{Fu2008,marra2016}.}\nocite{marra2016} The chemical
potential dependence is shown in Fig.~\ref{fig:phasediagram}(b),
together with the size of the $\chi=-1$ region around $\varphi=\pi$.
The behavior as a function of $\mu$ with constant $k_S$ exhibits
finite-size $k_FL$ oscillations from scattering at the NS interface,
which are not present \cite{sticlet2017,pientka2017-tsp} in the result
(not shown) for the matched case $\mu=\mu_{\mathrm{lead}}$, $m=m_{\mathrm{lead}}$ where
$\mu_*=\mathrm{const}(\mu)$. The correspondence between $\chi$
and the mode spectrum at $k_x=0$ is shown in
Fig.~\ref{fig:phasediagram}(c). The above narrow-channel approximation
breaks down when $|k_F|L\gtrsim1$, and is applicable for the first
lobe. This limitation is also visible in Fig.~\ref{fig:phasediagram}(c),
where the narrow-channel approximation predicts a zero-energy crossing between
$0<\tilde{k}_FL<\pi$, whereas in the exact solution the system is in the nontrivial
state for the whole interval.
Nevertheless, states with $\chi=-1$ can be achieved also at
higher doping and mismatch, but in a narrower parameter region.

It is important to note that S/FI bilayers have restrictions on the
magnitude of the exchange field.  The S/FI bilayer energy spectrum
becomes gapless at $h>\Delta$. Moreover, thin S/FI bilayers at low
temperatures support a thermodynamically stable superconducting state
only below the limit $h<\Delta(T=0)/\sqrt{2}$
\cite{clogston1962,chandrasekhar1962,maki1964-pps1} above which a
first-order transition to normal state occurs at $T=0$.  As the
induced effective order parameter is
$\Delta_*\propto|\cos\frac{\varphi}{2}|$, a change of the 1D invariant
can however be achieved for any $h$ at phase differences close enough
to $\varphi=\pi$.

\begin{figure}
  \includegraphics{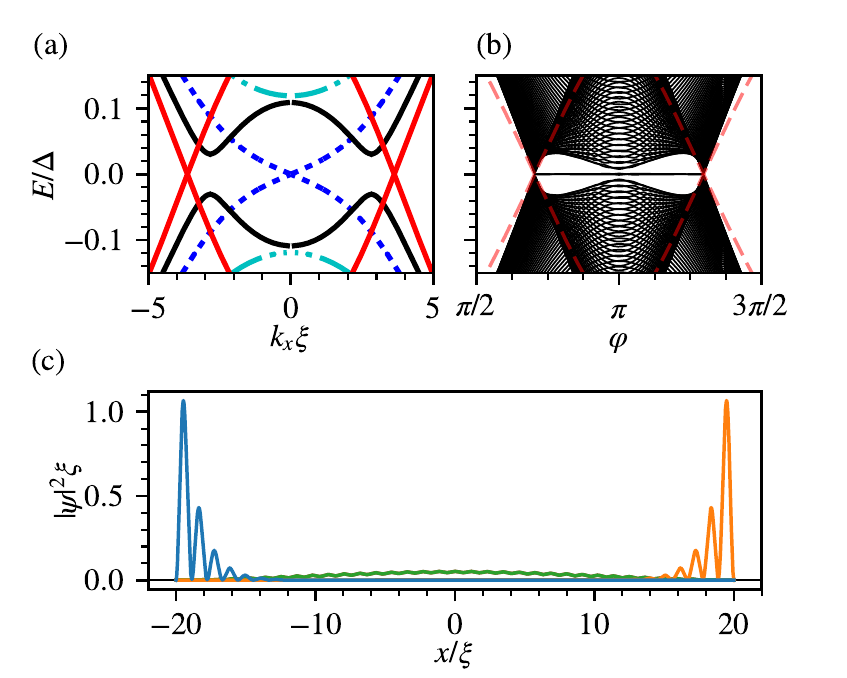}
  \caption{\label{fig:bound}
    (a) Mode dispersion in the
    trivial state
    $\varphi=0.6\pi$
    (dash-dotted),
    at gap closing
    $\varphi=\varphi_c=0.704\pi$
    (dotted), non-trivial state $\varphi=0.8\pi$ (black), and second gap closing $\varphi=\pi$ (red).
    Here
    $\mu=15\Delta$ and other parameters are as in
    Fig.~\ref{fig:phasediagram}.
    (b)
    Bound state spectrum as a
    function of $\varphi=\varphi_j$ for system size $\ell=40\xi$.
    The $E_g(k_x=0)$ bulk gap~\eqref{eq:Egk0} is also shown (red dashed).
    (c)
    Bound state wave function probability densities for
    $\varphi=0.8\pi$. Shown is the $E\approx0$ MBS (localized at the
    ends), and the lowest-energy extended
    state.  }
\end{figure}

The propagating mode spectrum of the 1D narrow-channel model
is shown in Fig.~\ref{fig:bound}(a) for different values of the  phase difference.
The behavior is typical to quantum wires \cite{kitaev2001-umf,oreg2010-hla}:  
the magnetic and superconducting
proximity effects open energy gaps at $k_x=0$ and $k_x=k_F$. The
energy gap at $k_x=0$
\begin{align}
  \label{eq:Egk0}
  E_g(k_x=0)\simeq{}Z_*\Big ||\vec{h}_*|-\sqrt{|\Delta_*|^2+\mu_*^2}\Big |
  \,,
\end{align}
closes and reopens at the topological transition. 
The gap at $k_F$
closes at $\varphi=\pi$ where $\Delta_*$ vanishes. 

\emph{Finite size.} The bound state energies
of a system with finite size in the $x$-direction
are given by the zeros of the
determinant of the effective 1D model,
\cite{gelfand1960-ifs,forman1987-fdg,waxman1994}
\begin{align}
  \label{eq:wfunc}
  w(\epsilon)
  =
  \Det[\epsilon - \mathcal{H}_{\rm eff}(\epsilon)]
  =
  \det\Bigl[
  \begin{pmatrix}1 & 0 \\ 0 & 0\end{pmatrix}
  +
  \begin{pmatrix}0 & 0 \\ 1 & 0\end{pmatrix}
  \Psi(\epsilon)
  \Bigr]
  \,.
\end{align}
Here  $\Psi(\epsilon)$ is the fundamental matrix connecting the ends of
the 1D channel, for the 1D differential operator
$\mathcal{H}_{\rm eff}$, defined analogously to
Eq.~\eqref{eq:transfermatrix} above.  Roots $w(\epsilon_j)=0$ of the
above $8\times8$ determinant can be found numerically.  The bound
state wave functions associated with each can be found from the
corresponding singular vectors.

Figure~\ref{fig:bound}(b) shows the bound state energy spectrum as a
function of the phase difference $\varphi$.  When the phase difference
crosses the bulk topological transition point, one of the ABS crosses
over to form a MBS pinned at $\epsilon\approx0$
and localized at the ends of the 1D channel [see
Fig.~\ref{fig:bound}(c)].

\begin{figure}
  \includegraphics{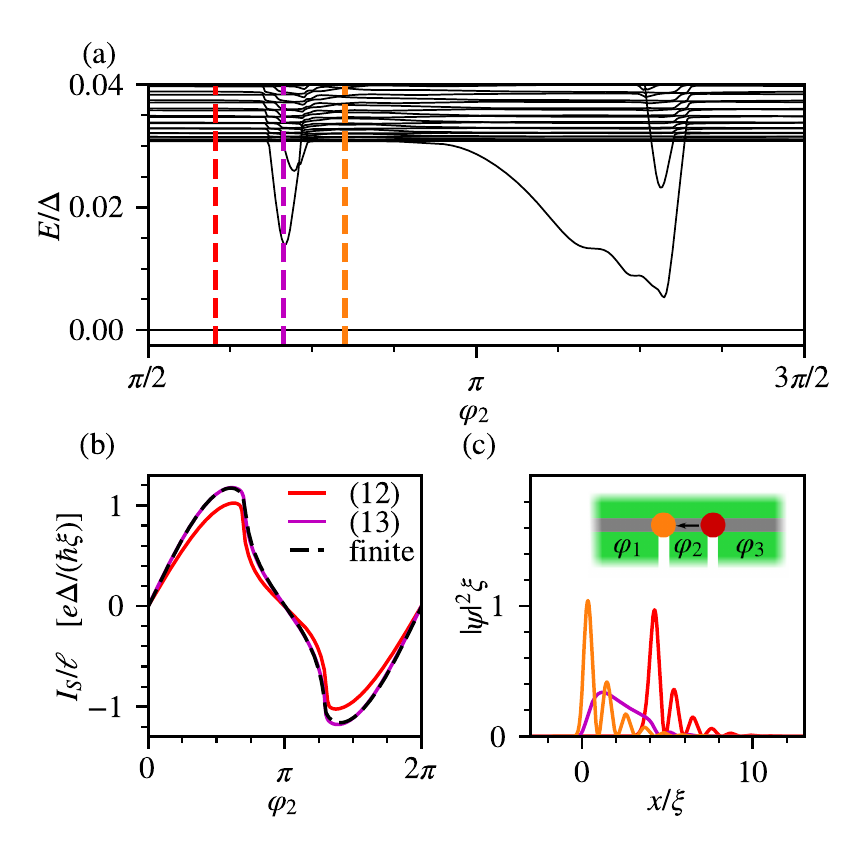}
  \caption{\label{fig:fig4}
    (a)
    Bound-state spectrum in a finite-size ``3-finger'' setup with finger widths
    $\ell_{j}=100\xi,4\xi,100\xi$ and phases $\varphi_j=0,\varphi_2,0.8\pi$,
    (b)
    Current density--phase relations, at $T=10^{-3}\Delta$.
    Shown is the result for uniform junction [Eq.~\eqref{eq:supercurrent-exact}],
    the same within the narrow-channel approximation [Eq.~\eqref{eq:supercurrent}],
    and $I_{S,2}/\ell_2$ in the 3-finger system.
    Parameters as in Fig.~\ref{fig:bound}.
    (c)
    Sweeping the phase $\varphi_2=0.6\pi\mapsto0.704\pi\mapsto0.8\pi$,
    the $\epsilon\approx0$ MBS localized at $x=\ell_2$ (red) moves toward $x=0$ (orange).
    Other parameters defined in caption of Fig.~\ref{fig:phasediagram}.
    Inset: schematic of the structure considered.
  }
\end{figure}

\emph{Supercurrent in a multiple-finger setup.}  Consider now the geometry of Fig.~\ref{fig:setup} with multiple superconducting fingers with widths
$\ell_j$ and different order parameter phases $\varphi_j$ on one side.
For a spatially piecewise constant Hamiltonian, we then have
$\Psi(\epsilon)=\prod_je^{W_j(\epsilon,\varphi_j)\ell_j}$.  Supercurrent
exiting the $j$th superconducting finger is given by the corresponding
derivative of the grand potential, which has a closed-form expression:
\begin{align}
  \label{eq:supercurrent-exact}
  I_{S,j}
  &= \frac{2e}{\hbar}T\partial_{\varphi_j}\ln\tr{}e^{-\beta H}
  =
  -\frac{e}{\hbar}T\sum_{\omega_n}\partial_{\varphi_j}\ln \Det\mathbf{M}(i\omega_n)
  \\
  \label{eq:supercurrent}
  &\simeq
  -\frac{e}{\hbar}T\sum_{\omega_n}\partial_{\varphi_j}\ln w(i\omega_n)
  \,.
\end{align}
The sum runs over the Matsubara frequencies
$\omega_n=2\pi{}T(n+\frac{1}{2})$.  Here, we used the fact that
starting from a functional determinant approach
\cite{gelfand1960-ifs,forman1987-fdg},
$\Det[i\omega-\mathcal{H}]\propto\Det\mathbf{M}(i\omega)\propto\Det[i\omega-\mathcal{H}_{\rm
  eff}]\propto{}w(i\omega)$, with overall proportionality constants
independent \cite{epaps} of $\varphi_j$.  Resulting current-phase
relations are shown in Fig.~\ref{fig:fig4}(b), for infinite-length
channels ($\ln\Det\mapsto\sum_{k_x}\ln\det$) and for a finite-length
$\ell<\infty$ system.  The results from \eqref{eq:supercurrent-exact}
and \eqref{eq:supercurrent} differ somewhat at $L=0.25\xi$, but
approach each other as the channel width $L\to0$. The finite-size
system result is close to the infinite-size result.

\emph{Phase control of the MBS.} Consider now a  superconducting finger of width $\ell_2$ between a trivial ($\varphi_1=0$, $x<0$) and a
nontrivial ($x>\ell_2$, $\varphi_3=0.8\pi$) segment.
The bound state spectrum as a function of $\varphi_2$ is shown in Fig.~\ref{fig:fig4}(a). 
As $\varphi_2$ crosses the bulk
transition point $\varphi_c\approx0.7\pi$ of the segment, the MBS
initially localized at $x=\ell_2$ re-localizes to $x=0$ (see Fig.~\ref{fig:fig4}(c)).
We have assumed $\ell_2\sim\xi$ so that the energy gap remains large also when sweeping
the phase.
The result shows that in a multi-finger setup, the MBS location can be controlled,
envisioning 2D channel networks with FI/S electrodes as a platform  for a
phase-controlled braiding of the MBS. \cite{Fu2008,hell2017-cbm}
To drive a segment into the non-trivial state ($\varphi\to\pi$)
  one can use a superconducting loop, connected to at least to some of the fingers.
  By controlling the supercurrent, it is likely possible to fine
    tune the MBS position.

InAs-2DEG\cite{Deon11,*Deontwo11,*Amado13,*Amadotwo2013,*Fornieri13}
with Al/EuS leads\cite{moodera1988electron,hao1990spin,Strambini2017}
provide a topological gap of $E_g/k_B\gtrsim\unit[60]{mK}$.  The
corresponding coherence length is $\xi\sim\unit[80]{nm}$, making the
fabrication of the devices compatible with the modern technologies.
Phase biasing can be implemented with superconducting loops and, in
combination with current injection, can be used to fine tune the MBS
position.

\emph{Conclusions.} In summary, we have used a transfer matrix
approach to obtain an effective 1D boundary Hamiltonian.
We have applied it to compute the spectrum of a S/FI--2DEG junction and to show
how the topological properties of the structure can be tuned by the
superconducting phase differences and the electrostatic gating.  This
enables the spatial control of the MBS, and
2D topological networks for braiding operations without requiring
strong external magnetic fields.
Our approach is quite general, not limited to any specific model or symmetry class, and can be
extended to other 1D channel problems with continuum Hamiltonians that
are polynomials in $k_y$. Moreover, the model can be applied to investigate the properties
  of Josephson junctions in two and three-dimensional systems,
  for example surfaces of topological insulators \cite{Fu2008,Kurter2015,amet2016-siq},
  or graphene \cite{titov2006-jeb} junctions.

\acknowledgments

P.V., E.S., A.B. and F.G. acknowledge funding by the European Research Council under the European Union's Seventh Framework Program (FP7/2007- 2013)/ERC Grant agreement No. 615187-COMANCHE. F.S.B acknowledges funding by the Spanish Ministerio de Econom\'ia y Competitividad (MINECO) (Projects No. FIS2014-55987-P and FIS2017-82804-P). A.B. acknowledges MIUR-FIRB 2012 RBFR1236VV and CNR-CONICET cooperation programme.

\nocite{bergeret2005-ots}

\clearpage

\onecolumngrid
\appendix

\setcounter{equation}{0}
\setcounter{figure}{0}
\renewcommand{\theequation}{S\arabic{equation}}
\renewcommand{\thefigure}{S\arabic{figure}}

\makeatletter
\let\origparagraph\paragraph
\renewcommand{\paragraph}[1]{\vskip 2ex\origparagraph{#1}}
\makeatother

\section{Intermediate steps taken}

As explained in the main text, we apply a scattering method
closely related to mode
matching approaches \cite{lee1981,hatsugai1993-esi,mora1985,garciamoliner1990-gtm}
and to computing functional determinants
\cite{gelfand1960-ifs,forman1987-fdg,waxman1994,kosztin1998-feo}.

\paragraph{Bound state equation.}
Let us point out the status of the bound state equation.  The result of
Refs.~\cite{forman1987-fdg} can be written as (note
that $\tr\mathbf{W}=0$ and $\det\mathbf{\Psi}=1$),
\begin{align}
  \Det G^{-1}(\epsilon)
  =
  \det\bigl[
  \mathbf{p}_- + \mathbf{p}_+ \mathbf{\Psi}(L/2,-L/2)
  \bigr]
  \,,
  \quad
  \mathbf{p}_-
  =
  \begin{pmatrix}
    1 & 0
    \\
    0 & 0
  \end{pmatrix}
  e^{-\mathbf{W}_-L_-}
  \,,
  \quad
  \mathbf{p}_+
  =
  \begin{pmatrix}
    0 & 0
    \\
    1 & 0
  \end{pmatrix}
  e^{\mathbf{W}_+L_+}
  \,,
\end{align}
where $L_\pm$ are the lengths of the upper$(+)$ and lower$(-)$
superconducting leads and $L$ that of the normal channel in between.
The multiplicative normalization of $\Det$ depends on the
highest-order derivative term in $\mathcal{H}$.

Consider the diagonalization
$\mathbf{W}_\pm=\mathbf{\Phi}_\pm\diag(\Lambda_\pm^{<},\Lambda_\pm^{>})\mathbf{\Phi}_\pm^{-1}$
with growing $(\Re\Lambda_\pm^>>0)$ and decaying
$(\Re\Lambda_\pm^<<0)$ modes, and write
$\mathbf{\Phi}=\begin{pmatrix}\phi^<&\phi^>\\w^<\phi^<&w^>\phi^>\end{pmatrix}$.
For $L_\pm\to\infty$ and neglecting modes in the leads that decay when
moving away from the N-region,
$\mathbf{p}_\pm\to\mathbf{R}\mathbf{P}_{\pm}$ where
$\mathbf{P}_{-}=\begin{pmatrix}1&-[w_-^>]^{-1}\\0&0\end{pmatrix}$,
$\mathbf{P}_{+}=\begin{pmatrix}0&0\\-w_+^<&1\end{pmatrix}$ are the
projection matrices in Eq.~(3) of the main text, and
$\mathbf{R}=\diag(R_-,R_+)$,
$R_-=\phi^{<}_-{}e^{-\Lambda_-^{<}{}L_-}(\phi_-^{<})^{-1}[1-(w_-^{>})^{-1}w_-^{<}]^{-1}$,
$R_+=\phi^{>}_+{}e^{\Lambda_+^{>}{}L_+}(\phi_+^{>})^{-1}[1-(w_+^{<})^{-1}w_+^{>}]^{-1}(-w_+^{<})^{-1}$,
depends only on the lead Hamiltonians. As a consequence,
$\Det G^{-1}(\epsilon)\simeq\Det{}R_+\Det{}R_-\Det\mathbf{M}_y$,
where $\Det{}R_\pm$ are independent of the phase of the order parameter.

\paragraph{Singular vectors.}
Each zero $\epsilon_j$ of the determinant is associated with one or more singular vectors
such that
$[\mathbf{P}_-+\mathbf{P}_+\Psi(L/2,-L/2)]\mathbf{u}_j=0$.
Each corresponds to a bound state wave function vector $\mathbf{u}(x)=\Psi(x,-L/2)\mathbf{u}_j$.

\paragraph{Green function.}
The Green function of the original Hamiltonian
at $|y|,|y'|<L/2$ can be expressed as
\begin{align}
  \label{eq:Gfunc}
  G(y,y')
  =
  \begin{pmatrix}1 & 0\end{pmatrix}
  \mathbf{\Psi}(y,y_0)
  [\mathbf{C}_+ \theta(y - y') - \mathbf{C}_- \theta(y' - y)]
  \mathbf{\Psi}(y_0,y')
  \begin{pmatrix}0 & 1\end{pmatrix}^T
  \,.
\end{align}
The constants $\mathbf{C}_\pm$ such that
$\mathbf{C}_++\mathbf{C}_-=\mathbf{1}$ are determined by boundary
conditions, and $y_0$ is an arbitrary fixed value.  Direct calculation
shows Eq.~\eqref{eq:Gfunc} then satisfies
$[\epsilon-\mathcal{H}]G(y,y')=\mathbf{1}\delta(y-y')$.
Matching to boundary conditions $\psi+A_\pm\hat{\partial}_y\psi=0$ and
setting $y_0=-L/2$, we have
$\mathbf{C}_+=\mathbf{M}_{y_0}^{-1}\mathbf{P}_-$,
$\mathbf{C}_-=\mathbf{M}_{y_0}^{-1}\mathbf{P}_+\mathbf{\Psi}(L/2,-L/2)$.

\paragraph{Local propagator.}
The diagonal resolvent is
$G(y,y)=\frac{1}{2}[G(y,y+0^+)+G(y+0^+,y)]=\frac{1}{2}\begin{pmatrix}1&0\end{pmatrix}\mathbf{q}(y)\begin{pmatrix}0&1\end{pmatrix}^T$.
From Eq.~\eqref{eq:Gfunc} and
$\mathbf{C}_+-\mathbf{C}_-=\mathbf{M}_{y_0}^{-1}\gamma_3\mathbf{M}_{y_0}$
we obtain Eq.~(5) in the main text.  Eq.~(6) follows after block
matrix algebra:
\begin{align}
  \mathbf{P}_-\mathbf{\Psi}(-\frac{L}{2},y)=\begin{pmatrix}p & q \\ 0 & 0\end{pmatrix}
  \,,
  \quad
  \mathbf{P}_+\mathbf{\Psi}(\frac{L}{2},y)=\begin{pmatrix}0 & 0 \\ r & s \end{pmatrix}
  \,,
  \quad
  G(y,y)^{-1}
  =
  q^{-1}p - s^{-1}r
  \,.
\end{align}

\paragraph{Free energy.}
The free energy can be expressed up to a constant as
$\mathcal{F}=-\frac{1}{2}T\sum_{\omega_n}\ln\Det{}G^{-1}(i\omega_n)$.
Since $\det{}\mathbf{R}$ and the normalization of $\Det$ are independent of the
order parameter phases,
$\mathcal{F}(\varphi)=-\frac{1}{2}T\sum_{\omega_n}\ln\Det{\mathbf{M}}(i\omega_n,\varphi)+\mathcal{F}_0$.
As a consequence, supercurrents
$I=\frac{2e}{\hbar}\partial_\varphi\mathcal{F}$ flowing in the
structure are determined only by $\ln\Det\mathbf{M}$, and by
extension, the phase-dependent part of its narrow-channel 1D
approximation,
$\mathcal{F}(\varphi)\simeq-\frac{1}{2}T\sum_{\omega_n}\ln\Det[\mathcal{H}_{\rm
  eff} - \epsilon]+\mathcal{F}_0'$.

\paragraph{Supercurrent.}
Note also that if restricting Eq.~(13) of the main text to low energies, and diagonalizing $\mathcal{H}_{\rm eff}$ yields a well-known result for the supercurrent
$I_{S,j}=-\frac{e}{2\hbar}\sum_{m}\tanh\bigl(\frac{\epsilon_m}{2T}\bigr)\partial_{\varphi_j}{\epsilon_m}$
where $\epsilon_m$ are the bound state energies.

\paragraph{Quasiclassical approximation.}
In the limit $\mu_{\rm lead}\to\infty$ and $\vec{\alpha}_{x/y}=0$ we
can compute the projectors $\tilde{\mathbf{P}}_\pm$.  Using an integral
representation of the matrix sign function, we have
\begin{align}
  \tilde{\mathbf{P}}_\pm
  =
  \int_{C_\mp}\frac{\dd{k_y}}{2\pi}
  \frac{1}{ik_y\mathbf{1} - \mathbf{W}_\pm(k_x)}
  =
  \int_{C_\mp}\frac{\dd{k_y}}{2\pi}
  \begin{pmatrix}
    ik_yG(k_x,k_y)b
    &
    G(k_x,k_y)
    \\
    -k_y^2 bG(k_x,k_y)b
    &
    ik_ybG(k_x,k_y)
  \end{pmatrix}
  \,,
\end{align}
where $b=\frac{\tau_3}{2m}$, and $C_{+(-)}$ are counter-clockwise
semicircles enclosing the upper(lower) complex half-plane. Poles
indicating propagating modes are displaced from the real axis by
$\Im\epsilon\ne0$.  Moreover,
$G(k_x,k_y)=[\epsilon - \mathcal{H}_{\rm lead}(k_x,k_y)]^{-1}$ is the Green
function.  Changing the integration
variable to $\xi=\frac{k^2}{2m_{\mathrm{lead}}} - \mu_{\mathrm{lead}}$
and taking the limit $\mu_{\mathrm{lead}}\to\infty$ we find
\begin{align}
  \tilde{\mathbf{P}}_\pm
  =
  \begin{pmatrix}
    \frac{1}{2} & \mp\frac{im_{\mathrm{lead}}}{k_{F,\mathrm{lead}}}\tau_3g_\pm
    \\
    \pm{}\frac{ik_{F,\mathrm{lead}}}{4m_{\mathrm{lead}}}g_\pm\tau_3 & \frac{1}{2}
  \end{pmatrix}
  \,,
\end{align}
where
$g_{+(-)}=\frac{i}{\pi}\fint_{-\infty}^\infty\dd{\xi}\tau_3G_{\mathrm{upper(lower)}}(\xi)$
are the quasiclassical functions in the leads, with
$\fint_{-\infty}^\infty\equiv\frac{1}{2}\int_{C_+}-\frac{1}{2}\int_{C_-}$. Since
$g_\pm^2=1$, reflecting the half-rank property of the projector
$\tilde{\mathbf{P}}_\pm$, the first and second rows of
$\tilde{\mathbf{P}}_\pm\mathbf{u}=0$ are the same equation.  Reduction to the
form discussed in the main text is then obtained by
$\tilde{\mathbf{P}}_\pm=\begin{pmatrix}\frac{1}{2}&mA_+\tau_3\\\frac{1}{4m}\tau_3A_-^{-1}&\frac{1}{2}\end{pmatrix}\mathbf{P}_\pm$.

\paragraph{Explicit quantum wire model.}
The order parameter $\hat{\Delta}_*$, the effective exchange field
$\vec{h}_*$, and the renormalization $Z_*$ read explicitly
($j=\mathrm{upper},\mathrm{lower}$)
\begin{subequations}
\begin{align}
  M_{j,\pm}
  &=
  \frac{D_j}{\sqrt{|\Delta_j|^2 - (h_j \mp \epsilon)^2}}
  \,,
  \qquad
  \hat{\Delta}_*
  =
  \frac{1}{2}
  \sum_{j,\pm}
  M_{j,\pm}
  [1 \pm \frac{\vec{h}_j}{h_j}\cdot\vec{\sigma}]
  [\Delta_j\tau_++\Delta_j^*\tau_-]
  \,,
  \\
  \vec{h}_*
  &=
  \sum_{j,\pm}
  M_{j,\pm}
  \frac{h_j \mp \epsilon}{2h_j}
  \vec{h}_j
  \,,
  \qquad
  Z_*^{-1}
  =
  1
  +
  \sum_{j,\pm}
  M_{j,\pm}
  \frac{\epsilon \mp h_j}{2\epsilon}
  \,.
\end{align}
\end{subequations}
Here, $D_j=\frac{\hbar^2k_S}{2mL}$ describes the effective strength of
the the coupling to the S leads, and $h_j=|\vec{h}_j|$.
The superconducting proximity effect
in the presence of the exchange field also induces an odd-frequency
triplet component in the pairing amplitude \cite{bergeret2005-ots}.
It is not important for the main physics here as the triplet component vanishes at $\epsilon\to0$.

\paragraph{A matrix identity.}
For piecewise constant Hamiltonians with fundamental matrix
$\Psi=\prod_{j=1}^n\Psi_j$, one has
$\det[M + N\Psi]=\det[M + N\Psi_n\cdots\Psi_1]$,
$\Psi_j=\Phi_je^{iK_j(x_{j+1}-x_j)}\Phi_j^{-1}$ where
$K_j=\diag(k_{j,1},\ldots,k_{j,m})$. For large $|\Im{}K||x_{j+1}-x_j|$, the
matrix product is numerically unstable to evaluate. The following
identity can be used to improve the conditioning:
\begin{gather}
  \det[M + N \Psi_n\ldots\Psi_1]
  =
   \frac{
     \det B
   }{
     \prod_j \det \Phi_j
   }
   \exp(\sum_j\tr Q_j)
   \,,
   \\
   B
   \equiv
   \begin{pmatrix}
     M\Phi_1(x_1) &             &              & &   N\Phi_n(x_{n+1})
     \\
     -\Phi_1(x_2) & \Phi_2(x_2) &              &  &   
     \\
                  & -\Phi_2(x_3) & \Phi_3(x_3) &  &   
     \\
                  &            & \ddots       & \ddots   &
     \\
                  &            &              &       -\Phi_{n-1}(x_n) & \Phi_{n}(x_n)
   \end{pmatrix}
   \,,
\end{gather}
where $\Phi_j(x)=\Phi_je^{i K_j (x-x_j) - Q_j}$ and $Q_j=\diag(q_{j,1},\ldots,q_{j,m})$
such that
\begin{align}
  q_{j,p}
  =
  \begin{cases}
    0 \,, & \text{for $\Im k_{j,p}\ge0$,}
    \\
    -(x_{j+1}-x_{j})\Im k_{j,p} \,, & \text{for $\Im k_{j,p}<0$.}
    \\
  \end{cases}
  \,.
\end{align}
The matrix $B$ is typically better conditioned and its log-det can be
evaluated via standard methods. Note that $B$ explicitly encodes the boundary
and wave function matching conditions.

\begin{figure}[b]
  \includegraphics{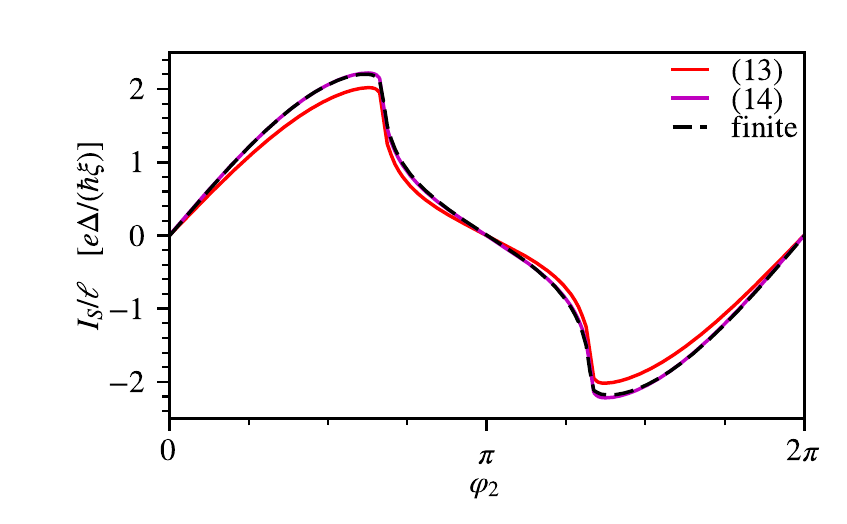}
  \caption{
    \label{fig:4b-short}
    Same as Fig. 4(b), but for $L=0.1\xi$.
  }
\end{figure}


\end{document}